\documentclass[a4paper]{article}

\usepackage{INTERSPEECH2022}

\usepackage{wasysym}
\usepackage{hyperref}       %
\usepackage{booktabs}       %
\usepackage{amsfonts}       %
\usepackage{amsmath}
\usepackage{array}
\usepackage{graphicx}
\usepackage{multirow}
\usepackage{xcolor}
\usepackage{cite}

\newcommand{\Patts}{$\mathcal{C}_1$}
\newcommand{\Legion}{$\mathcal{C}_2$}
\newcommand{\Hol}{Sp\textsubscript{1}}
\newcommand{\Jmp}{Sp\textsubscript{2}}

\newcommand{\accentspeaker}{Sp\textsubscript{accent}}
\newcommand{\etal}{\textit{et al.}}
\newcommand{\eg}{\textit{e.g.,}}

\title{Training Text-To-Speech Systems From Synthetic Data: \\ A Practical Approach For Accent Transfer Tasks}
\name{Lev~Finkelstein,~~Heiga~Zen,~~Norman~Casagrande$^\dagger$,~~Chun-an~Chan,~~Ye~Jia,~~Tom~Kenter, \\ Alexey~Petelin,~~Jonathan~Shen\sthanks{~~Work done while at Google.},~~Vincent~Wan,~~Yu~Zhang,~~Yonghui~Wu,~~Rob~Clark}

\address{Google LLC,~~$^\dagger$DeepMind }
\email{finklev@google.com}
\begin{document}

\maketitle

\begin{abstract}
  Transfer tasks in text-to-speech (TTS) synthesis --- where one or more aspects of the speech of one set of speakers is transferred to another set of speakers
  that do not feature these aspects originally --- remains a challenging task.
  One of the challenges is that models that have high-quality transfer capabilities can have issues in stability,
  making them impractical for user-facing critical tasks.
  This paper demonstrates that transfer can be obtained by training a robust TTS system on data generated by a less robust TTS system designed for a high-quality transfer task;
  in particular, a CHiVE-BERT monolingual TTS system is trained on the output of a Tacotron model designed for accent transfer.
  While some quality loss is inevitable with this approach, experimental results show that the models trained on synthetic data this way can produce high quality audio displaying accent transfer, while preserving speaker characteristics such as speaking style.
\end{abstract}

\section{Introduction}\label{section:introduction}
Transfer tasks in TTS, where certain aspects of the speech of one or more speakers are transferred to other speakers whose speech does not originally have these aspects, include multiple tasks, such as style transfer \cite{wang2018style,zhanglearning}, transfer of vocal qualities \cite{jia2018transfer,hsu2018hierarchical} and prosody transfer \cite{skerry2018towards,klimkov2019fine,lee2019robust}.

The accent transfer task \cite{hsu2018hierarchical,zhang2019learning} aims to generate speech in accent, \eg British English for a North American speaker.
This task can be approached from multiple directions, such as cascading a TTS system and voice conversion-based accent conversion \cite{zhao2018accent} or making a TTS system to be able to synthesize arbitrary combination of speakers and accents via disentangling speakers and accents \cite{hsu2018hierarchical}.
In this paper we build on the latter approach; we use a Tacotron-based model \cite{zhang2019learning}, designed to disentangle speaker and language, and apply it to the accent transfer task, where it transfers the accent of speakers in the training data to a target speaker, who is also present in the training data, but who is not associated with the accent.

Tacotron-based TTS systems are known to offer high-quality TTS.
Sometimes, however, they exhibit stability issues \cite{liu2020teacher,shen2020non}, such as early stopping, word skipping, word repetition, and babbling.
These issues makes Tacotron-based TTS less attractive for user-facing critical applications, such as voice-based assistants.
To make use of the transfer capabilities of Tacotron without running into the issues described above, we propose a two-step approach where, first, a large amount of speech data is synthesized by a Tacotron-based accent transfer model \cite{zhang2019learning}, and subsequently, a more robust model, which is by itself not capable of the transfer task, is trained on both natural recordings of speakers in their native accent, and on speech synthesized by the Tacotron in the target accents.
We use the CHiVE-BERT prosody model~\cite{kenter2020chivebert} together with a WaveGRU synthesizer model~\cite{wavernn,wavegru}, as the stable second model.

We demonstrate that the synthetic speech-based knowledge transfer network yields solid performance in terms of naturalness and accent transfer capability, while preserving speaker characteristics, including speaking style.
Evaluations show that the quality loss inherently associated with the switch to synthetic data is within acceptable bounds: up to 0.3 in 5-scale mean opinion score (MOS).

\section{Related Work}\label{section:related}

The use of synthetic data has been explored before to improve the quality of TTS systems.
Sharma \etal~\cite{sharma2020strawnet} proposed \emph{StrawNet} in which an autoregressive (AR) WaveNet~\cite{oord2016wavenet} model is used to generate high quality synthetic training data which is then used to train another AR WaveNet (teacher) and non-AR Parallel WaveNet (student) \cite{oord2017parallel} models.
Similarly Hwang \etal~\cite{hwang2021tts} proposed an effective method for training a non-AR TTS model (FastSpeech 2 \cite{ren2020fastspeech}) when the amount of training data is insufficient.
The work adds synthetic speech corpora generated from a well-designed AR TTS model (Tacotron \cite{okamoto2019tacotron}) to their training data.
These papers focused on improving the quality of non-AR TTS in low-data regime, whereas this paper discusses the use of synthetic data for the accent transfer task.

In the style transfer task, Ribeiro \etal~\cite{ribeiro2022cross} proposed cross-speaker style transfer for TTS using voice conversion.
A voice conversion model was used to convert expressive speech by source speakers to those by a target speaker.
In addition to the original speech in the neutral speaking style by the target speaker, the converted expressive speech was used to train a speaker-dependent expressive TTS system for the target speaker.
This work used voice conversion to synthesize expressive-style data with the target speaker's voice characteristics, while the proposed approach uses a Tacotron-style TTS model with accent transfer to produce speech in the target accent with the target speaker's voice characteristics.

The use of synthetic data from the first model to train the second model is similar to knowledge distillation \cite{hinton2015distilling}, since the second model is trained to predict outputs from the first model.

\section{Using synthetic data for accent transfer}\label{section:methodology}

\begin{figure}[t]
  \centering
  \includegraphics[width=\linewidth]{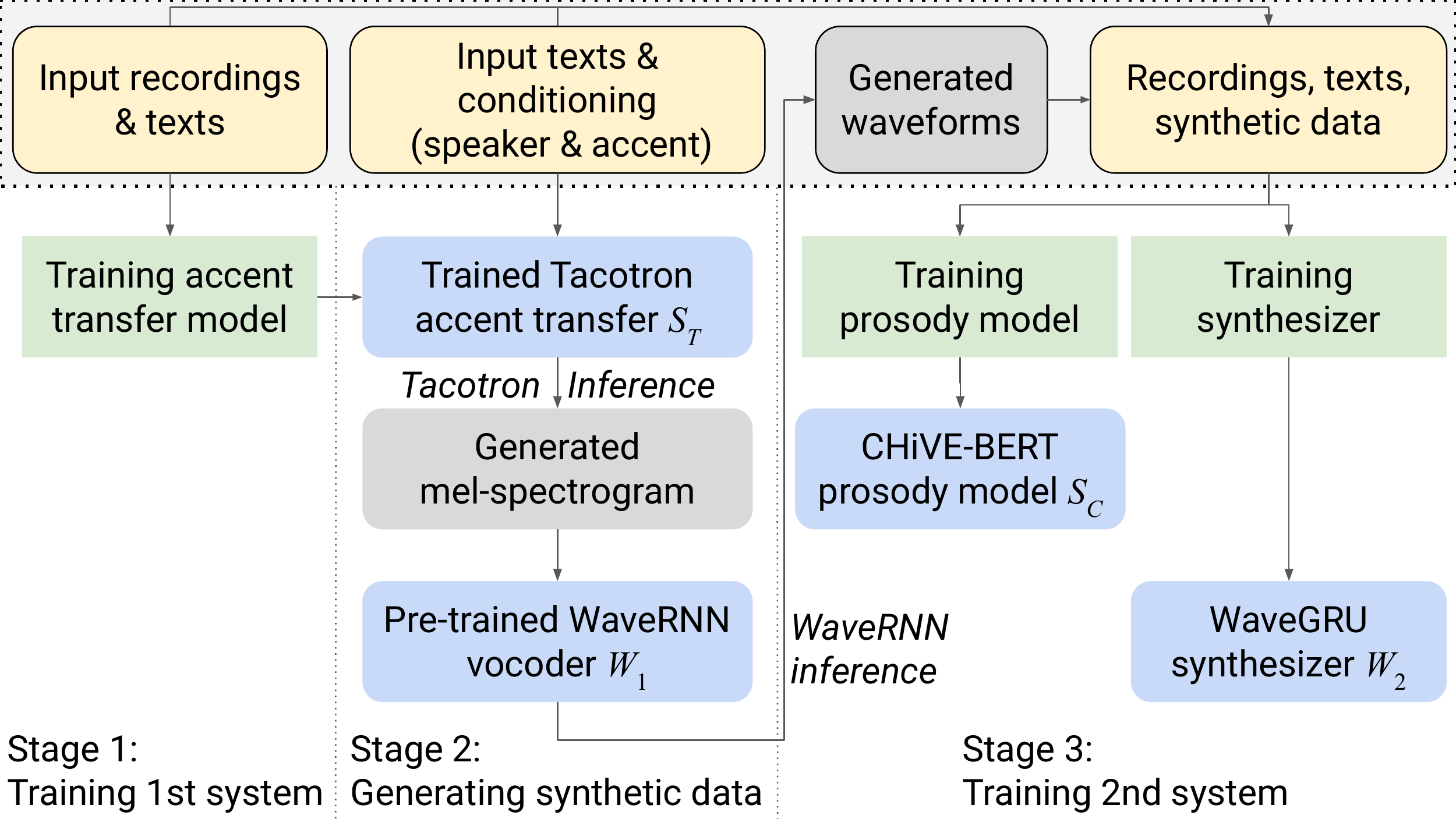}
  \caption{A schematic illustration of the proposed two-level training approach.}
  \label{fig:two_level_scheme}
\end{figure}

This section describes how training on synthetic data works for the accent transfer task.
A CHiVE-BERT system~\cite{kenter2020chivebert} is trained on synthetic data generated by a Tacotron-based accent transfer system.
The framework illustrated in Figure~\ref{fig:two_level_scheme} involves the following four components:
\begin{enumerate}
\item A Tacotron-based accent transfer system $S_T$ that generates mel-spectrograms given a phoneme sequence \cite{zhang2019learning}.
\item A pretrained WaveRNN neural vocoder $W_1$ that transforms mel-spectrograms to waveforms \cite{wavernn}.
\item A CHiVE-BERT system $S_C$ that generates prosodic features (durations, fundamental frequency values) given linguistic features \cite{kenter2020chivebert}.
Although its accent transfer performance is limited, it is reliable in performing a standard monolingual TTS task.
\item A dense version of single-band WaveGRU synthesizer $W_2$ with a simplified gated recurrent unit (GRU) cell~\cite{cho2014properties} and logistic mixture output \cite{wavegru}.
This system takes linguistic and prosodic features as its input then synthesizes speech waveforms.
\end{enumerate}

See Section~\ref{section:discussion} for further discussion of each component mentioned above.
The details of the training process used in the setup, as well as the experimental results, are presented in the next section.

\newcommand{\femalespeaker}[1]{FS\textsubscript{#1}}  %
\newcommand{\malespeaker}[1]{MS\textsubscript{#1}} %

\section{Experimentation}\label{section:experimentation}
This section describes the experimental results for the two-level accent transfer training task.
Audio samples are available at {\footnotesize \url{https://google.github.io/chive-prosody/chive-bert-synthetic/}}.

\subsection{Experimental Setup}\label{subsection:setup}
We used two proprietary read speech corpora\footnote{The corpora meet
the Google AI Principles~\url{https://ai.google/principles/}.}, \Patts\ and \Legion.
\Patts\ contains English speech, read in a natural style, by 58 speakers
with different accents:
255K utterances by 34 speakers in North American accent (\textsf{US}),
54K utterances by 5 speakers in Australian accent (\textsf{AU}),
60K utterances by 8 speakers in British accent (\textsf{GB}),
34K utterances by 7 speakers in Indian accent (\textsf{IN}),
and 10K utterances by 4 speakers in other accents.

Corpus \Legion\ contains 12K utterances by 40 speakers in \textsf{US}.
It was read in either neutral or news reading style, which allows us to test if characteristics such as style are preserved while accent is being transferred.
The goal is to perform accent transfer to \textsf{GB}, \textsf{AU}, and \textsf{IN} for eight specific speakers in \Legion\ that are mostly associated with the news style.
Note that the two corpora have no overlap in terms of speakers, i.e., there are no speakers who have multiple accents, though individual speakers have their own native accents.
As a result of the way data and speakers are distributed across corpora \Patts\ and \Legion, we end up with \textsf{GB}, \textsf{AU}, and \textsf{IN} news reading style speakers, while we only have neutral training data for these speakers, and we only have news reading style data for \textsf{US} speakers.

In what follows, \Hol\ and \Jmp\ represent the \textsf{US} female and \textsf{US} male speaker respectively that have most data in the \Patts\ corpus.
Also from \Patts, we have \accentspeaker\ for each accent, which is the speaker with most data in that accent.

The training process consisted of the following steps:
\begin{enumerate}
\item \textbf{Training the Tacotron model}: The Tacotron accent transfer model followed~\cite{zhang2019learning} without adversarial loss.\footnote{Preliminary experiments showed that the adversarial loss was not essential if there were enough speakers per accent in the training data.}
We further modified the global VAE to a hierarchical version~\cite{rosenberg2019speech} which improved the stability of our teacher Tacotron model.
The model was trained on all data of \Patts\ and \Legion\ combined.
As \textsf{US} had more data than other accents, we upsampled the data of the speakers in \textsf{AU}, \textsf{GB}, and \textsf{IN} until each of them had the same number of utterances as the largest speaker in \textsf{US}.

\item \textbf{Synthetic data generation}: For each target accent, we synthesized data with the Tacotron model for eight selected speakers in \Legion, and for \Hol\ and \Jmp.
We used transcripts for the same eight \Legion\ speakers and regarded their style as news reading style.
We also synthesized speech for the transcripts of \accentspeaker, which we regarded as neutral style.
These style labels were later used while training a CHiVE-BERT model.
Note that this labeling was done only to reflect the implicit conditioning on the text.
\item \textbf{Training the second system}. The CHiVE-BERT-based system was trained, per accent, on all the synthetic data above.
  The corresponding WaveGRU synthesizer model was trained, per accent, on the same synthetic data as the CHiVE-BERT system, plus the original recordings of \accentspeaker.
\end{enumerate}
Note that this setup is only one of the possible configurations.
Some choices are explained in more detail in Section~\ref{section:discussion}.

Lastly, a natural alternative approach would be to train a multi-accent CHiVE-BERT prosody model plus a WaveGRU synthesizer directly on the same accent transfer setup, like the Tacotron model described above.
However, preliminary experiments showed that the speech produced by such a system had low naturalness and transfer capability.
Thus we did not pursue this direction any further.

\subsection{Evaluation Methodology}

The trained systems (i.e., \{$S_C$, $W_2$\} pairs) were evaluated by human raters on \textsf{GB}, \textsf{AU}, and \textsf{IN} for \Hol, \Jmp, and eight \Legion\ speakers mentioned in the synthetic data generation in Section~\ref{subsection:setup}.
In addition, we conducted evaluations for the same voices without accent transfer.
Subjective mean opinion score (MOS) \cite{MOS} style evaluations were performed as follows:
\begin{itemize}
\item \textbf{Naturalness (short-form)}: To focus on just accent transfer, only short sentences were used in this setup, as there was relatively little difference between the neutral and news reading style for short sentences.
\item \textbf{Accent quality}: A special kind of appropriateness test, where raters were asked to evaluate if the voice sample belonged to a native accent speaker.
\item \textbf{Naturalness (long-form)}: Same as above, but longer sentences were used, to focus on the naturalness of the accent-transferred voices while the news reading style was employed.
\item \textbf{News-reading style appropriateness}: Appropriateness of the voice sample as a news reader, on entire news paragraphs.
\end{itemize}

The ratings were 1 to 5 (1: bad, 2: poor, 3: fair, 4: good, and 5: excellent).
Human speech typically gets an average score of 4.4 or higher, though this highly depends on the specific task.
Note that rater pools and human perception of style appropriateness and accent transfer could differ among the locales, thus the absolute numbers per locale are not directly comparable.
The news-reading style appropriateness test may be considered the most important one, as performance can be affected by accent, naturalness, or both.

\subsection{Experimental Results}

In this section we present major results on two male and two female speakers from \Legion\ that we denote \femalespeaker{i} and \malespeaker{i}.

\subsubsection{Quality of accent transfer voices built on synthetic data}
In this main experiment, we tested two aspects of the voice quality: naturalness and accent quality tests.
The results are shown in Table~\ref{tab:naturalness-appropriateness}. We also
present the data for the same speakers in \textsf{US} generated with the system
described in~\cite{kenter2020chivebert}.
It can be seen from the table that the accent transfer quality is reasonably good (MOS score of 4 or higher).

\begin{table}[t]
  \centering
  \caption{Accent quality and style appropriateness MOS results for accent transfer voices trained on synthetic data.}
  \begin{tabular}{*{4}{c}}
    \toprule
    \multirow{2}{*}{\textbf{Accent}} &\multirow{2}{*}{\textbf{Speaker}} & \multirow{2}{*}{\textbf{Accent quality}} & \textbf{Naturalness}\\
                                     &                                  &                         & \textbf{(short-form)}\\\midrule

    \multirow{4}{*}{\textsf{GB}} &
     \femalespeaker{1} & $4.05 \pm 0.07$ & $4.19 \pm 0.05$\\
    &\femalespeaker{2} & $4.08 \pm 0.07$ & $4.12 \pm 0.05$\\
    &\malespeaker{1}   & $4.39 \pm 0.05$ & $4.31 \pm 0.04$\\
    &\malespeaker{2}   & $4.40 \pm 0.05$ & $4.31 \pm 0.04$ \\\midrule

    \multirow{4}{*}{\textsf{AU}} &
     \femalespeaker{1} & $4.28 \pm 0.04$ & $4.38 \pm 0.08$\\
    &\femalespeaker{2} & $4.42 \pm 0.05$ & $4.46 \pm 0.04$\\
    &\malespeaker{1}   & $4.39 \pm 0.04$ & $4.42 \pm 0.05$\\
    &\malespeaker{2}   & $4.41 \pm 0.05$ & $4.39 \pm 0.05$ \\\midrule

    \multirow{4}{*}{\textsf{IN}} &
     \femalespeaker{1} & $4.44 \pm 0.06$ & $4.11 \pm 0.07$\\
    &\femalespeaker{2} & $4.29 \pm 0.08$ & $4.21 \pm 0.06$\\
    &\malespeaker{1}   & $4.45 \pm 0.06$ & $4.22 \pm 0.06$\\
    &\malespeaker{2}   & $4.32 \pm 0.06$ & $4.21 \pm 0.06$ \\\midrule

   \multirow{4}{*}{
     \begin{tabular}{@{}cc@{}}
      \textsf{US} \\ (reference) \\
     \end{tabular}} &
     \femalespeaker{1} & -- & $4.22 \pm 0.06$\\
    &\femalespeaker{2} & -- & $4.22 \pm 0.06$\\
    &\malespeaker{1}   & -- & $4.06 \pm 0.06$\\
    &\malespeaker{2}   & -- & $4.21 \pm 0.06$ \\
    \bottomrule
  \end{tabular}
  \label{tab:naturalness-appropriateness}
\end{table}

Although the naturalness of the synthetic voices rated close to the naturalness of the reference voices, a proper cross-accent comparison is not feasible as it is mentioned above.

\subsubsection{Style transfer quality}
We also validated if speaker characteristics were kept during the accent transfer process.
In particular, we checked this by analyzing news-style reading appropriateness and long-form naturalness.
The results are shown in Table~\ref{tab:news-style}, and the MOS scores
are perceived as good as well.
If we compare Table~\ref{tab:naturalness-appropriateness} and Table~\ref{tab:news-style}, we see that the short-form and long-form naturalness scores are close for most of the speakers in most cases.
\begin{table}[t]
  \centering
  \caption{Naturalness MOS results for accent transfer voices trained on synthetic data.}
  \begin{tabular}{*{4}{c}}
    \toprule
    \multirow{2}{*}{\textbf{Accent}} &\multirow{2}{*}{\textbf{Speaker}} & \textbf{Style} & \textbf{Naturalness}\\
                                     &                                  & \textbf{appropriateness} & \textbf{(long-form)}\\\midrule

    \multirow{4}{*}{\textsf{GB}} &
     \femalespeaker{1} & $4.11 \pm 0.05$ & $4.15 \pm 0.07$\\
    &\femalespeaker{2} & $3.99 \pm 0.07$ & $4.12 \pm 0.06$\\
    &\malespeaker{1}   & $4.26 \pm 0.05$ & $4.32 \pm 0.05$\\
    &\malespeaker{2}   & $4.27 \pm 0.06$ & $4.25 \pm 0.06$ \\\midrule

    \multirow{4}{*}{\textsf{AU}} &
     \femalespeaker{1} & $4.24 \pm 0.04$ & $4.26 \pm 0.06$\\
    &\femalespeaker{2} & $4.03 \pm 0.06$ & $4.41 \pm 0.05$\\
    &\malespeaker{1}   & $4.32 \pm 0.04$ & $4.36 \pm 0.06$\\
    &\malespeaker{2}   & $4.14 \pm 0.04$ & $4.15 \pm 0.07$ \\\midrule

    \multirow{4}{*}{\textsf{IN}} &
     \femalespeaker{1} & $4.29 \pm 0.06$ & $4.16 \pm 0.06$\\
    &\femalespeaker{2} & $4.37 \pm 0.06$ & $4.25 \pm 0.07$\\
    &\malespeaker{1}   & $4.35 \pm 0.06$ & $4.34 \pm 0.05$\\
    &\malespeaker{2}   & $4.49 \pm 0.05$ & $4.35 \pm 0.06$ \\
    \bottomrule
  \end{tabular}
  \label{tab:news-style}
\end{table}

\subsubsection{Quality loss from intermediate (Tacotron) to final (CHiVE-BERT) model}
One natural question is what loss in quality there is between the first Tacotron-based model and the final CHiVE-BERT model.
The comparison of accent transfer quality and style appropriateness between the
intermediate Tacotron model and the final CHiVE-BERT model are shown in
Tables~\ref{tab:accent-taco-vs-cadenza} and~\ref{tab:appropriateness-accent}
respectively.
As expected, training from synthetic data leads to some loss in quality.
However, the quality of the speech yielded by the final system is still high (above 4.0 in MOS).

It can also be seen from the table that accent quality of the final model was significantly affected by the intermediate model.
For example, accent quality of the final model for female speakers in \textsf{GB} was significantly worse than for other speakers.
This seems to be an artifact of the intermediate Tacotron model; accent quality of the intermediate model for female speakers in \textsf{GB} was also significantly worse than other for speakers.

\begin{table}[t]
  \centering
  \caption{Accent transfer quality of the intermediate (Tacotron) and the final (CHiVE-BERT) models.}
  \begin{tabular}{*{5}{c}}
    \toprule
    \multirow{2}{*}{\textbf{Accent}} & \multirow{2}{*}{\textbf{Speaker}} & \textbf{Intermediate} & \textbf{Final} & \multirow{2}{*}{\textbf{Diff.}} \\
    &  & \textbf{model} & \textbf{model} & \\\midrule

    \multirow{4}{*}{\textsf{GB}} &
     \femalespeaker{1} & $4.34 \pm 0.06$ & $4.05 \pm 0.07$ & 0.29 \\
    &\femalespeaker{2} & $4.23 \pm 0.07$ & $4.08 \pm 0.07$ & 0.15 \\
    &\malespeaker{1}   & $4.46 \pm 0.05$ & $4.39 \pm 0.05$ & 0.07 \\
    &\malespeaker{2}   & $4.54 \pm 0.04$ & $4.40 \pm 0.05$ & 0.14 \\\midrule

    \multirow{4}{*}{\textsf{AU}} &
     \femalespeaker{1} & $4.48 \pm 0.04$ & $4.28 \pm 0.04$ & 0.20 \\
    &\femalespeaker{2} & $4.59 \pm 0.03$ & $4.42 \pm 0.05$ & 0.17 \\
    &\malespeaker{1}   & $4.57 \pm 0.04$ & $4.39 \pm 0.04$ & 0.18 \\
    &\malespeaker{2}   & $4.40 \pm 0.04$ & $4.41 \pm 0.05$ & 0.01 \\\midrule

    \multirow{4}{*}{\textsf{IN}} &
     \femalespeaker{1} & $4.60 \pm 0.05$ & $4.44 \pm 0.06$ & 0.16\\
    &\femalespeaker{2} & $4.59 \pm 0.05$ & $4.29 \pm 0.08$ & 0.30\\
    &\malespeaker{1}   & $4.60 \pm 0.04$ & $4.45 \pm 0.06$ & 0.15\\
    &\malespeaker{2}   & $4.50 \pm 0.05$ & $4.32 \pm 0.06$ & 0.18\\
    \bottomrule
  \end{tabular}
  \label{tab:accent-taco-vs-cadenza}
\end{table}

\begin{table}[t]
  \centering
  \caption{Appropriateness of the intermediate model (Tacotron) vs. the final model (CHiVE-BERT).}
    \begin{tabular}{*{5}{c}}
      \toprule
      \multirow{2}{*}{\textbf{Accent}} & \multirow{2}{*}{\textbf{Speaker}} & \textbf{Intermediate} & \textbf{Final} & \multirow{2}{*}{\textbf{Diff.}} \\
      &  & \textbf{model} & \textbf{model} & \\\midrule

    \multirow{4}{*}{\textsf{GB}} &
     \femalespeaker{1} & $4.45 \pm 0.05$ & $4.11 \pm 0.05$ & 0.34 \\
    &\femalespeaker{2} & $4.41 \pm 0.05$ & $3.99 \pm 0.07$ & 0.42 \\
    &\malespeaker{1}   & $4.44 \pm 0.06$ & $4.26 \pm 0.05$ & 0.18 \\
    &\malespeaker{2}   & $4.47 \pm 0.05$ & $4.27 \pm 0.06$ & 0.20 \\ \midrule

    \multirow{4}{*}{\textsf{AU}} &
     \femalespeaker{1} & $4.48 \pm 0.04$ & $4.24 \pm 0.04$ & 0.24 \\
    &\femalespeaker{2} & $4.59 \pm 0.03$ & $4.03 \pm 0.06$ & 0.56 \\
    &\malespeaker{1}   & $4.57 \pm 0.04$ & $4.32 \pm 0.04$ & 0.25 \\
    &\malespeaker{2}   & $4.40 \pm 0.04$ & $4.14 \pm 0.04$ & 0.26 \\ \midrule

    \multirow{4}{*}{\textsf{IN}} &
     \femalespeaker{1} & $4.49 \pm 0.06$ & $4.29 \pm 0.06$ & 0.20 \\
    &\femalespeaker{2} & $4.56 \pm 0.05$ & $4.37 \pm 0.06$ & 0.19 \\
    &\malespeaker{1}   & $4.52 \pm 0.06$ & $4.35 \pm 0.06$ & 0.17 \\
    &\malespeaker{2}   & $4.52 \pm 0.05$ & $4.49 \pm 0.05$ & 0.03 \\
    \bottomrule
  \end{tabular}
  \label{tab:appropriateness-accent}
\end{table}

\subsubsection{Other observations}
One interesting observation we made is that, when trained on synthetic data, the loss was much lower than when human recordings are used.
We hypothesize that human recordings have a higher variance than the synthetic data, and are thus more difficult to learn.
Similar observations were reported in \cite{hwang2021tts}.

Another observation was that the speakers could sound differently across accents, \eg\ generated \textsf{AU} speech was faster than the \textsf{GB} speech.
The reason is that different accents are associated with differences in, for example, speed and pitch, and this is reflected in the training data.
Further analysis is out of the scope of the current paper since the relevant changes occur at the level of the accent transfer Tacotron system.

The last thing to note is that using a WaveGRU synthesizer \cite{wavegru} in the final model performed significantly better than using a WaveRNN \cite{wavernn} model,
which yielded multiple artifacts\footnote{The artifacts were severe enough to exclude these systems from being subjectively evaluated.}.
One hypothesis is that if the waveform generation models in the first and the second system are of the same architecture,
they both might be liable to same bias. As such, the second system may learn to emulate the bias of the first system,
instead of performing a more general learning task, leading to deteriorated performance.

\section{Discussion}\label{section:discussion}
The scheme of training from synthetic data as shown in
Figure~\ref{fig:two_level_scheme} required a careful design of the training scheme.
This section describes some of the most important considerations.

\textbf{Choice of the vocoder $W_1$}: The neural vocoder should produce as few artifacts as possible.
Even if the average human listener cannot perceive these artifacts in the synthetic speech, they may be captured and can potentially accumulate in the second model since the synthetic data is used to train the second system.
As data generation is an offline process, it is better to use the best performing neural vocoder for $W_1$.

\textbf{Choice of the synthesizer $W_2$}: It is not guaranteed that the synthesizer model trained from original recordings of human speech produces sufficiently high-quality audio as $W_2$.
When we used a WaveGRU synthesizer trained on the human recordings only, the resulting audio contained too many artifacts, possibly due to mismatch between the training and the inference stage.
These artifacts disappeared after retraining the synthesizer on a combination of human recordings and synthetic data.

\textbf{Choice of the amount of the synthetic data}: Preliminary experiments showed that too little synthetic data resulted in quality degradation, mostly on the style axis.\footnote{We don't present subjective scores in this section
since these evaluations were done on the initial tuning stage, and the system has changed since that.
However, due to the nature of the observations, we believe that the same observations should be valid for any system of this type.}
However, although there are no limits on the amount of the synthetic data that can be generated per speaker,
the quality improvement saturated around 50k--60k utterances per speaker.

\textbf{Choice of the nature of the synthetic data}: For style transfer, it can be beneficial to use transcripts that are associated with the desired style.
In our experiments we observed that the CHiVE-BERT model trained on data synthesized from the transcripts of \accentspeaker\ sounded more natural, while being less appropriate in terms of style, than the same model trained on the data synthesized from all the transcripts from \Legion.
The mixture of the two sources provided a good tradeoff between naturalness and style appropriateness scores.

\textbf{Choice of the synthetic data vs. human recordings}:
When a transform task contains multiple parameters (\eg\ accent and style), it is important to keep the balance between original recordings and synthetic data.
In our experiments, adding human recordings to the training data of the synthesizer $W_2$ significantly improved the audio fidelity.
However, it had no impact on the quality of the CHiVE-BERT model $S_C$.
Since the CHiVE-BERT model and the WaveGRU synthesizer have different roles (prosody is modeled by CHiVE-BERT, whereas phonetics, voice quality and identity by the synthesizer), we used different training data for them.

\section{Conclusions}\label{section:conclusions}

We presented a methodology for performing accent transfer by training a monolingual CHiVE-BERT system along with a WaveGRU synthesizer on synthetic data generated by a Tacotron-based accent transfer model.
The approach can be useful when the Tacotron accent transfer system is inapplicable for some reason -- in our case, due to reliability issues.

The experiments show that we can get a reliable system that produces high quality audio in all dimensions --
naturalness, accent transfer quality, and style appropriateness.
We also analyzed quality loss while comparing to the primary Tacotron system as
an upper quality bound for this task, and showed that this loss is acceptable.

While the experiments were conducted on an accent transfer task, the set of experiments that measure quality loss while training from synthetic data are task-independent.
We would expect therefore that a similar loss would appear in other tasks involving training from synthetic data, and that the given methodology
is not limited to the accent transfer task, but can be used for
virtually every task that includes an almost-perfect/experimental TTS system and a
reliable TTS system lacking specific features.
Each such specific application
may require proper tuning
as discussed in Section~\ref{section:discussion}.

In future work, we plan to investigate the applicability of our approach on other TTS transfer tasks.

\appendix

\newpage
\bibliographystyle{IEEEtran}
\bibliography{bibliography}

\end{document}